\documentclass[a4paper]{article}
\usepackage{blindtext, graphicx}
\usepackage{authblk}

\usepackage[font=small]{caption}

\usepackage{url}

\usepackage{comment}

\usepackage[utf8]{inputenc}

\begin{document}


\providecommand{\keywords}[1]
{
  \small	
  \textbf{Keywords---} #1
}

\title{Performance vs Programming Effort between Rust and C on Multicore Architectures: Case Study in N-Body

}

\author[1]{Manuel Costanzo}
\author[1]{Enzo Rucci}
\author[1]{Marcelo Naiouf}
\author[1]{Armando De Giusti}



\affil[1]{III-LIDI, Facultad de Inform\'atica,
Universidad Nacional de La Plata - CIC\\
La Plata, Buenos Aires, Argentina\\
 \authorcr
\{mcostanzo,erucci,mnaiouf,degiusti\}@lidi.info.unlp.edu.ar}

\date{{July 20, 2021}}

\maketitle

\begin{center}
\texttt{This is the accepted version of the manuscript that was sent to review to \textit{2021 Latin American Computing Conference (CLEI)}. This manuscript was finally accepted for publication on July 20th, 2021 will be available online at IEEExplore Digital Library.}
\end{center}

\begin{center}
\texttt{\textregistered 2021 IEEE. Personal use of this material is permitted.  Permission from IEEE must be obtained for all other uses, in any current or future media, including reprinting/republishing this material for advertising or promotional purposes, creating new collective works, for resale or redistribution to servers or lists, or reuse of any copyrighted component of this work in other works.}
\end{center}

\clearpage

\begin{abstract}

Historically, Fortran and C have been the default programming languages in High-Performance Computing (HPC). In both, programmers have primitives and functions available that allow manipulating system memory and interacting directly with the underlying hardware, resulting in efficient code in both response times and resource use. 
On the other hand, it is a real challenge to generate code that is maintainable and scalable over time in these types of languages. In 2010, Rust emerged as a new programming language designed for concurrent and secure applications, which adopts features of procedural, object-oriented and functional languages. Among its design principles, Rust is aimed at matching C in terms of efficiency, but with increased code security and productivity. This paper presents a comparative study between C and Rust in terms of performance and programming effort, selecting as a case study the simulation of N computational bodies (N-Body), a popular problem in the HPC community. Based on the experimental work, it was possible to establish that Rust is a language that reduces programming effort while maintaining acceptable performance levels, meaning that it is a possible alternative to C for HPC.
\end{abstract}


\section{Introduction}

HPC is the use of extraordinary computing power systems and parallel processing techniques to solve complex problems with high computational demand~\cite{hpc_def}. Achieving this purpose requires having not only architectures that provide the necessary processing capacity, but also software that allows the problem to be efficiently computed. This is why choosing a programming language is not a trivial choice, and its selection will have an impact on both application performance and the required programming effort.

In general, HPC systems must compute the problem efficiently in order to improve program response times. To achieve this, the base language must provide primitives and functions that take advantage of the underlying hardware. This involves being able to generate multiple parallel tasks, that are capable of synchronizing and communicating with each other, and that take advantage of data locality and processor vector units, among other features.

Currently, the most popular languages in the HPC field are Fortran and C, which are characterized as languages with a low level of abstraction~\footnote{Usually simply referred to as \textit{low-level}.}. This class of languages allows the programmer to have an exhaustive control of the program and the data it processes, which can significantly improve response times as well as the use of application resources~\cite{chr12}. In turn, there are libraries that allow extending base language functionality and provide concurrent and parallel processing capabilities for multiprocessor architectures both with shared and distributed memory~\cite{openmp08, graham06, gropp02}.

Despite being widely used, it is a real challenge to generate code that is maintainable and scalable over time in these languages. This is why, in recent years, many programming languages with a high level of abstraction~\footnote{Also known simply as \textit{high-level}.} have tried to add support for concurrency and parallelism, in an attempt to compete with C and Fortran. Among these, we can mention Java~~\cite{gui13, sha14} and Python~\cite{mas11, jan20}; however, unfortunately, neither has been successful at becoming an alternative in the HPC community for the time being.

In 2010, Mozilla released Rust~\cite{kla18}, which is a language designed for the development of highly concurrent and secure systems. Rust is a compiled language that performs security checks on memory accesses at compile time~\cite{fred19}. It also supports direct hardware access and fine-grained control over memory representations, allowing algorithms to be specifically optimized~\cite{jung17}. Among its design principles, Rust seeks to provide a system control equivalent to C, but without losing sight of memory accesses security~\cite{bala17}, high performance, and code readability and scalability\cite{joha20}.

In recent years, Rust has increased its popularity and use in different fields to such an extent that in 2019 it was recognized as the most \textit{loved} programming language for the fourth consecutive year on Stack Overflow~\cite{ruststack}. Likewise, several companies continue to migrate their systems (or part of them) to Rust, among which the following stand out: Facebook~\cite{edenscm}, Mozilla~\cite{servo}, Discord~\cite{discord}, Dropbox~\cite{dropbox}, NPM~\cite{npm}, and others~\cite{rustcompanies}.

As previously mentioned, knowing the advantages and disadvantages of each programming language to implement HPC codes is essential, as well as being familiar with their statements and directives to obtain high-performance applications. In order to contribute to the evaluation of Rust as an alternative base language for parallel processing applications, this paper focuses on its comparison with C in terms of performance and programming effort. For this, N-Body – a problem with high computational demand that is popular in the HPC community – is selected as case study. The contributions of this work are as follows:

\begin{itemize}
    \item An optimized implementation in the Rust language that computes N-Body on multicore architectures, which is available in a public web repository for the benefit of the academic and industrial communities~\footnote{\url{https://github.com/ManuelCostanzo/Gravitational_N_Bodies_Rust}}.
    \item A rigorous comparison of Rust and C codes for N-Body in a multicore architecture considering performance and programming effort. Through this comparative study, we hope to help identify Rust's strengths and weaknesses for use in HPC.
\end{itemize}

This article is organized as follows: in Section~\ref{background}, the theoretical framework is introduced. Section~\ref{implementation} details the C algorithm used and the N-Body optimizations developed in Rust, from initial implementation to final version. Then, in Section~\ref{resexp}, the results obtained through the tests carried out are presented and discussed. Next, in Section~\ref{trabrel}, some related works are mentioned, and finally, in Section~\ref{conclusiones}, our conclusions and possible future work are presented.

\section{Background}\label{background}

\subsection{Programming languages for HPC}\label{lengprog}

Starting in the 1950s, programming languages developed very rapidly, which resulted in the creation of hundres of them. The challenge focuses both on the creation of new languages that provide novel tools aimed at reducing response times and programming cost, as well as on the adaptation of existing languages, improving them as technologies envolve~\cite{langcomparacion}.

Even though there is a wide variety of languages, their features will be more efficient for certain applications. When choosing a programming language for HPC, focus is usually on the impact it will have on the system. For this reason, these languages should provide tools that enable an efficient development of the solution. In this sense, the language should offer at least the following features:

\begin{itemize}
    \item Pointers that allow directly manipulating memory contents.
    \item Set of bit manipulation operators which, when used properly, can improve program times considerably.
    \item Support for different flags or compilation options that optimize code based on support architecture.
    \item Ability to use/embed native underlying architecture instructions to take advantage of its hardware characteristics.
    \item And others.
\end{itemize}

In addition to this, in order to improve response times, the language should offer tools or libraries that allow extending base language functionality to provide concurrent and parallel processing capabilities for multiprocessor architectures, both with shared (f.e., OpenMP~\cite{openmp08}) and distributed memory (f.e., OpenMPI~\cite{graham06} or MPICH~\cite{gropp02}).

\subsection{Rust}\label{lengRust}

Rust is a programming language that came to light in 2010 in the Mozilla company as a personal project of Graydon Hoare, who was in search of a language that would allow writing extremely fast code, at the same level as C, but without the memory management problems that historically led to errors in accesses and data races~\footnote{The data race is a condition that occurs when 2 or more threads access a shared/global variable and at least one of the threads writes it.}. As an added goal, the language should also offer the possibility of developing code using elements from procedural programming languages together with some from object-oriented and functional programming languages.

\subsubsection{Main Features}\label{rustcaracteristicas}

Rust's main features include the no need for a \textit{Garbage Collector} (freeing memory is rarely necessary), the concept of \textit{ownership}, and the concept of \textit{borrowing}.

The ownership model is based on 3 main rules:

\begin{enumerate}
    \item Every value in Rust has a variable called \textit{owner}.
    \item There can only be one owner at a time.
    \item When the owner goes out of the scope of the application, the value is removed.
\end{enumerate}

This concept has been gaining popularity both in the scientific community and among leading language developers, and is a unique feature of Rust~\cite{jung17}.

Other important concepts are those of reference and borrowing. Reference allows obtaining a value without taking ownership of it, while borrowing is a consequence of the reference: when functions have references as parameters, there is no need to return ownership because they never had it. Reference rules are as follows:

\begin{itemize}
    \item At any time, you can have one mutable reference or any number of immutable references.
    \item References must always be valid.
\end{itemize}

\subsubsection{Concurrency and parallelism}\label{rustconcurrencia}

Rust was designed with the focus on concurrency. The concept of ownership not only eliminates common errors in memory, but also avoids many race condition errors. In C, these errors are usually difficult to detect, because they occur under specific circumstances that are difficult to reproduce. Thanks to the concepts of ownership and borrowing, Rust detects many of these errors (but not all) at compile time, making it a very suitable language for problems that require concurrency and parallelism~\cite{kla18}.

Rust has a specific library for data parallelism, called Rayon~\cite{rayon}. Rayon seeks to convert Rust sequential code into parallel code in a simple way, guaranteeing the absence of data race and ensuring that all threads will see the updated value through what are called mutable references~\cite{rustconcurrency}.

The central mechanism by which the library defines the tasks that can run in parallel is \textit{join}. Rayon works in a similar way to OpenMP parallel regions, but instead of always executing the selected code in parallel, it dynamically decides if it makes sense to run it this way based on system resources at the time~\cite{rayonparallelism}.

Additionally, Rayon provides iterators that go through data structures and perform actions in parallel under the \textit{work stealing} technique, which consists in allowing a thread that has finished to ``steal'' tasks from another thread, thus avoiding idle threads and helping finish work earlier.

\subsection{The Gravitational N-Body Problem}\label{nbody}

This problem consists in simulating the evolution of a system composed by N bodies during a time-lapse. Each body presents an initial state, given by its speed and position. The motion of the system is simulated through discrete instants of time. In each of them, every body experiences an acceleration that arises from the gravitational attraction of the rest, which affects its state.

Newtonian mechanics support the simulation basis~\cite{Tipler2004}. The simulation is performed in 3 spatial dimensions and the gravitational attraction between two bodies $C_i$ and $C_j$ is computed using Newton's law of universal gravitation:

\[F=\frac{G \times m_i \times m_j}{r^2 }\]

where $F$ corresponds to the \textit{magnitude} of the gravitational force between bodies, $G$ corresponds to the gravitational constant~\footnote{Equivalent to  $6.674 \times 10^{11}$}, $m_i$ corresponds to the body mass of $C_i$, $m_j$ corresponds to the body mass of $C_j$, and $r$ corresponds to the Euclidean distance~\footnote{The Euclidean distance is given by 
$\sqrt{((x_j-x_j)^2+(y_j - y_i)^2+(z_j - z_i)^2)}$, where ($x_i$, $y_i$, $z_i$) is the position of $Ci$ and ($x_j$, $y_j$, $z_j$) is the position of $C_j$.}  between $C_i$ and $C_j$.

When $N$ is greater than 2, the gravitational force on a body corresponds to the sum of all gravitational forces exerted by the remaining $N-1$ bodies. The force of attraction leads each body to accelerate and move, according to the Newton's second law, which is given by the following equation:

\[F=m \times a\]

where $F$ is the force vector, calculated using the magnitude obtained from the equation of gravitation and the direction and sense of the vector going from the affected body to the body exerting the attraction.

Regarding the above equation, it is clear that the acceleration of a body can be calculated by dividing the total force by its mass. During a small time interval $dt$, the acceleration $a_i$ of the body $C_i$ is approximately constant, so the change in velocity is approximately:

\[dv_i=a_i dt\]

The change in body position is the integral of its speed and acceleration over the $dt$ time interval, which is approximately: 

\[dp_i=v_i dt + \frac{a_i}{2} dt^2 = (v_i + \frac{dv_i}{2})dt \]

In this formula, it is clear that one half of the position change is because of the old speed while the other half is because of the new speed. This integration scheme is called Leapfrog~\cite{leapfrog}.

\begin{figure}[htbp]
    \centering
    \includegraphics[width=7.5cm]{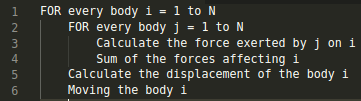}
    \caption{Pseudocode of the N-Body problem implementation}
    \label{ncuerpospseudo}
\end{figure}

The pseudo-code of the direct solution is shown in Fig.~\ref{ncuerpospseudo}. This problem presents two data dependencies that can be noted in the pseudo-code. First, one body cannot move until the rest have finished calculating their interactions. Second, they cannot advance either to the next step until the others have completed the current step.

\section{Implementation}\label{implementation}

\subsection{C Implementation}

To make a fair comparison between C and Rust, an algorithm optimized for multicore architectures presented in~\cite{rucci20} was used. The algorithm has the following features:

\begin{itemize}
    \item Multi-threading: through OpenMP directives, performs parallel force calculations separately from the one that moves the bodies. For this, use the \textit{for} directive with the \textit{schedule(static)} clause, to distribute the number of bodies evenly among the threads.
    \item Scalar optimizations: Optimizations are performed to reduce the computational cost of simple operations that are executed several times
     \begin{itemize}
        \item To perform a power, use the function 
        \textit{powf} when the data type is float and \textit{pow} when the data type is double. 
        \item Specify constants as floating where applicable.
        \item Replace divisions that have a constant as denominator by multiplications with the multiplicative inverse of that constant.
        \end{itemize}
    \item Vectorization: to vectorize, use the OpenMP 4.0 \textit{simd} directive that forces the vectorization of the code.
    \item Block processing: in order to exploit data locality, bodies are processed using a blocking technique. To achieve this, bodies are divided in fixed-size portions called \textit{blocks}. The bodies loop  is replaced by two others: a loop that iterates over all blocks and an inner loop that iterates over the bodies of each block.
    \item Loop unrolling: is an optimization technique that can improve the performance of a program by unrolling loop iterations into individual operations. This was done using the  \textit{unroll} directive from the compiler.
\end{itemize}

\begin{figure}[htbp]
    \centering
    \includegraphics[width=7.5cm]{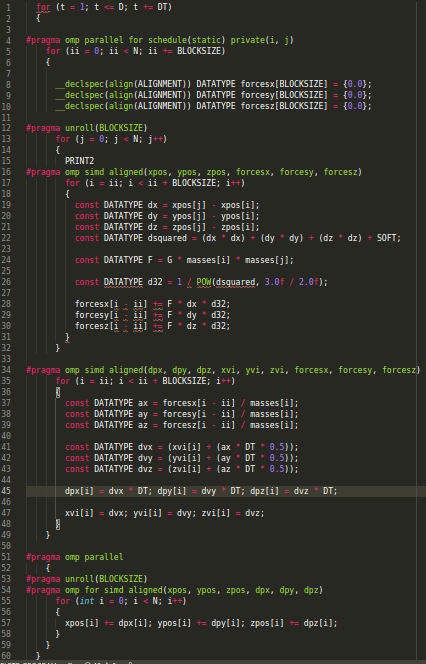}
    \caption{A snippet of the optimized C implementation}
    \label{c_code}
\end{figure}

Fig~\ref{c_code} shows a snippet of the C code used.

\subsection{Rust Implementation}

\subsubsection{Baseline code}

\begin{figure}[htbp]
    \centering
    \includegraphics[width=7.5cm]{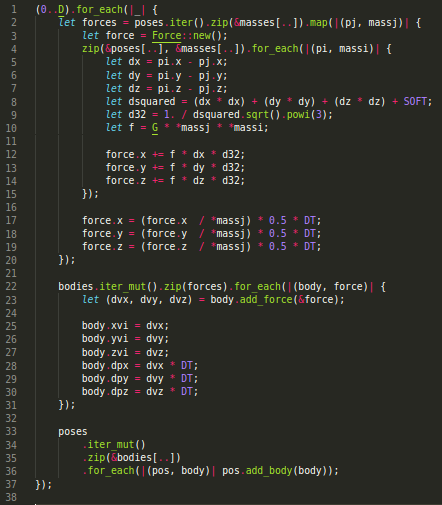}
    \caption{Rust sequential implementation}
    \label{fig:secuencial}
\end{figure}

Initially, a sequential version was developed, which is shown in Fig~\ref{fig:secuencial}. The first \textit{for\_each} (line 1) iterates over the total steps in time of the simulation. Then, between lines 2 and 20, an iterator is created that calculates the forces for each of the bodies. Within it, there is a loop (line 4), which iterates through each position to calculate the new force. In line 4, a \textit{zip}~\footnote{An iterator that iterates two other iterators simultaneously.} is performed that concatenates the mass for each position and calculates the new force of each body. Finally, in the iterator in line 22, the bodies are shifted.

Starting with this version, different optimizations that have been applied incrementally were considered.

\subsubsection{Multi-threading}

In this version, data processing is parallelized through the use of the \textit{Rayon} library. To achieve this goal, the following problem dependencies must be observed: (1) forces must be fully calculated before bodies can be moved; and (2) all threads must have finished moving their bodies before those new positions can be used for calculating the forces for the next iteration.

In this sense, Fig~\ref{fig:secuencial} shows the sections of the code where synchronization is required: as already explained, in line 2 the iterator that calculates the forces is created. In line 22, for each body the new force is concatenated, where the force of the body is updated (lines 23 to 30), fulfilling dependency 1. Finally, in line 33, the bodies are repositioned to later move on to the next iteration ($D+1$), respecting dependency 2.

\begin{figure}[htbp]
    \centering
    \includegraphics[width=7.5cm]{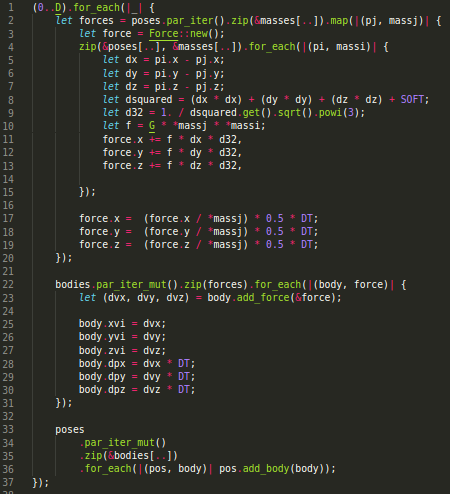}
    \caption{Parallel Rust Implementation}
    \label{fig:paralelo}
\end{figure}

Fig~\ref{fig:paralelo} shows that iterators can be parallelized just adding the \textit{par\_} prefix to them.  In \textit{Rayon}, each parallel section adds an implicit barrier, ensuring thread synchronization.

\subsubsection{\textit{Fold} iterator}
\begin{figure}[htbp]
    \centering
    \includegraphics[width=7.5cm]{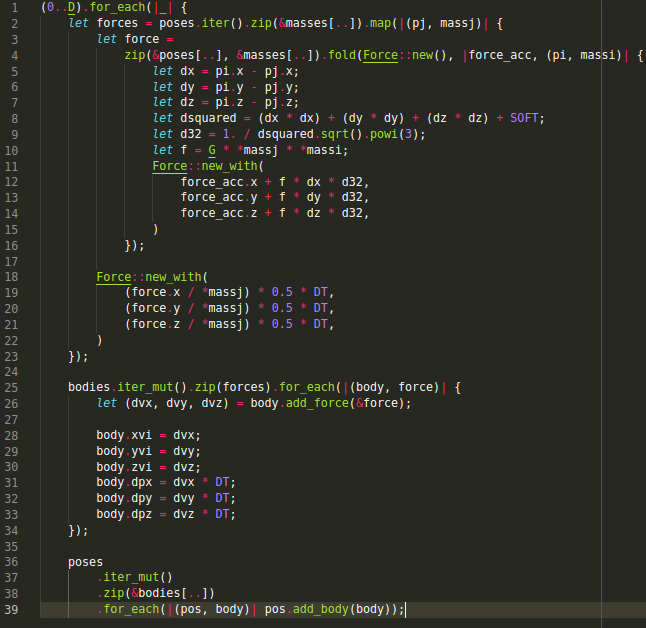}
    \caption{Parallel Rust implementation with \textit{fold}}
    \label{fig:paralela2}
\end{figure}

In line 4 of Fig~\ref{fig:paralelo}, iteration through each position is required to calculate the new force. In short, since the result of this iteration will be a single piece of data, it can be optimized using the \textit{fold} method (\textit{reduce}, in functional programming), which will update the force in each iteration. The modified code including this method can be seen in Fig~\ref{fig:paralela2}, starting from line 3.

\subsubsection{Mathematical optimizations}
Rust provides intrinsics~\cite{rustintrinsics} that allow applying optimizations in mathematical operations based on algebraic rules, which in turn allows computing with greater speed, but sacrificing precision. Even though they are not yet in the stable version of Rust, these intrinsics are widely used to improve the performance of mathematical operations. Among these, the following are of note:

\begin{itemize}
    \item \textit{fadd\_fast}: optimized sum.
    \item \textit{fsub\_fast}: optimized subtraction.
    \item \textit{fmul\_fast}: optimized multiplication.
    \item \textit{fdiv\_fast}: optimized division.
    \item \textit{frem\_fast}: optimized remainder.
\end{itemize}

To make use of these instructions, the \texttt{fast-floats} library~\footnote{This library is located at https://github.com/bluss/fast-floats and the extension https://github.com/bluss/fast-floats/pull/2 was used to avoid non-definite results.} was used, which implements them transparently and does not need to call the  \texttt{*\_fast} routines in each operation.

\subsubsection{Vectorization}

Vectorization allows performing the same operation on multiple data in parallel, being one of the most common techniques to improve the response times in an application. Through directives that are sent to the LLVM compiler, it is possible to generate specific vectorized code for the target platform, instead of the basic vector instructions.



By default, Rust tries to vectorize code automatically, but if code complexity prevents the compiler from doing this on its own, the programmer can generate the vectorized code
manually. In this case, manual vectorization was not required for the algorithm developed (the binary code for this was verified). As the the processor used for this work supports
AVX-512 instructions, the flag \texttt{target-feature=+avx512f} was sent to the compiler, to generate vectorized code of this type.

\subsubsection{Jemalloc}

By default, all programs in Rust use the system allocator, which is general purpose. Fortunately, there are allocators that are better adapted depending on the type of application, \textit{Jemallocator}~\cite{jemallocator} being one of the most used allocators for concurrent applications.

The main benefit of this allocator is scalability in multiprocessor and multithreaded systems. It does this by using multiple arenas, which are raw memory fragments from which allocations are made. Jemalloc works as follows: in multi-threaded environments, Jemalloc creates many arenas (four times more arenas than processors), and threads are mapped to these arenas in a circular fashion. This reduces the dealys between threads, since they will only compete if they belong to the same arena; otherwise, the threads will not affect each other. In turn, Jemalloc tries to optimize cache locality to have contiguous access to the data in memory~\cite{jemalloc}.

\begin{figure}[htbp]
    \centering
    \includegraphics[width=7.5cm]{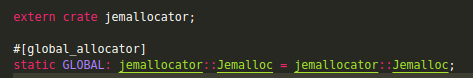}
    \caption{Jemalloc configuration in Rust code}
    \label{fig:jemallocatorconfig}
\end{figure}

To use Jemalloc in Rust, the Jemallocator library should be imported and the system allocator configured as shown in Fig~\ref{fig:jemallocatorconfig}.

\subsubsection{Block processing}

In order to take advantage of data locality, computing for this problem can be restructured to perform block processing. This leads to changing data access logic and where operations are carried out, among other changes. Additionally, for each architecture, both block size and the number of threads to execute will have to be configured.

\begin{figure}[htbp]
    \centering
    \includegraphics[width=7.5cm]{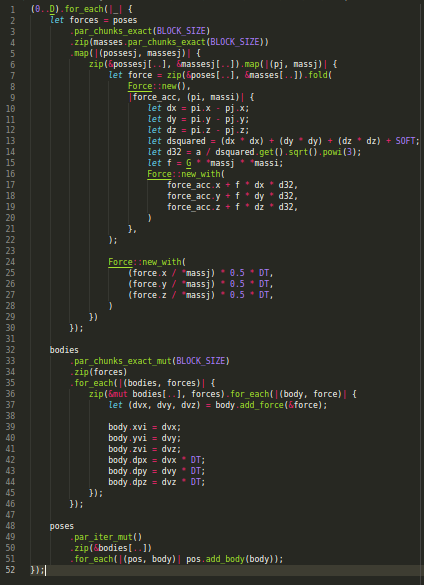}
    \caption{Rust parallel implementation with block processing}
    \label{fig:bloques}
\end{figure}

In this work, an attempt was made to process blocks of varying sizes: 8, 16, and 32. The code in Fig~\ref{fig:bloques} shows the algorithm processing by blocks. The central idea is that each process works with blocks of \textit{BLOCK\_SIZE} size, attempting to exploit the data locality present in this problem.

\section{Experimental Results}\label{resexp}

\subsection{Experimental Design}
All tests were performed on a 2$\times$Intel Xeon Platinum 8276 server with 28-core 2.20Ghz (with AVX-512 instructions), 256 GB of RAM and 112 hardware threads. To compile the code in C, the ICC compiler (version 19.1.0.166) was used, sending the following compilation flags:
    \begin{itemize}
        \item \texttt{-fp-model fast=2}: it enables math approximations.
        \item \texttt{-xCORE-AVX512}: it enables AVX-512 instructions.
        \item \texttt{-qopt-zmm-usage}: encourages the compiler to use the SIMD instructions.
        \item \texttt{-qopenmp}: OpenMP library..
    \end{itemize}
As for Rust, version 1.48.0 of the language and version 11.0 of the LLVM compiler were used. The compilation flag \texttt{-Ctarget-feature=+avx512f} was sent to enable AVX-512 vector instructions.

For the tests carried out, firstly, a base version of the Rust algorithm was used and the impact of applying each optimization incrementally was analyzed~\footnote{The previous version is labeled as \textit{Reference} in all figures.}. Then, the best version of Rust was selected to compare with its C equivalent. Tests were split into single-precision and double-precision tests, varying both workload (N = \{65536, 131072, 262144, 524288, 1048576\}) and number of threads (T = \{56, 112, auto\}) and block size (BLOCK\_SIZE = \{8, 16, 32\}) where necessary.

The programs worked on the same input data, after verifying that they produce the same outputs.

\subsection{Performance}

The GFLOPS (billion FLOPS) metric is used to evaluate performance.

\[GFLOPS=\frac{20 \times N^2 \times I}{T \times 10^9 }\]

where \textit{N} is the number of bodies, \textit{I} is the number of steps, \textit{T} is execution time (in seconds), and factor 20 represents the number of floating point operations required for each iteration.

\begin{figure}[htbp]
    \centering
    \includegraphics[width=0.95\columnwidth]{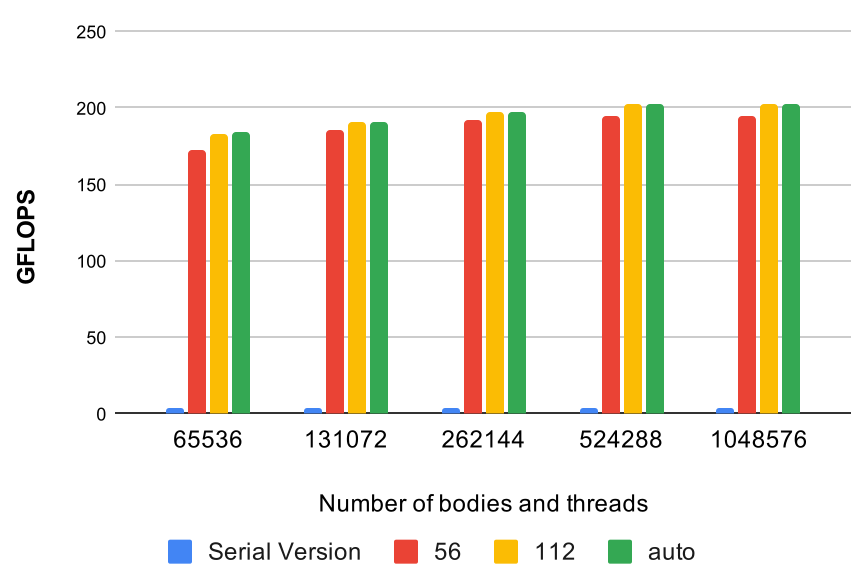}
    \caption{Performances obtained for different numbers of bodies and threads in the first parallel version}
    \label{res_multihilado}
\end{figure}

Fig~\ref{res_multihilado} shows the performances obtained when varying the number of bodies and the number of threads in the simplest parallel version (without any type of optimization), together with the times of the sequential version. The latter hardly reaches 4 GFLOPS. With multi-threading enabled, the parallel version is up to $50\times$ faster than the serial one . 
On the other hand, it can be seen that the use of \textit{hyper-threading} provides a small performance gain (up to 5\%). Finally, it can also be noted that Rust's \texttt{auto} option properly selects the number of threads to use, achieving FLOPS rates similar to those obtained with manual setting.

\begin{figure}[htbp]
    \centering
    \includegraphics[width=0.95\columnwidth]{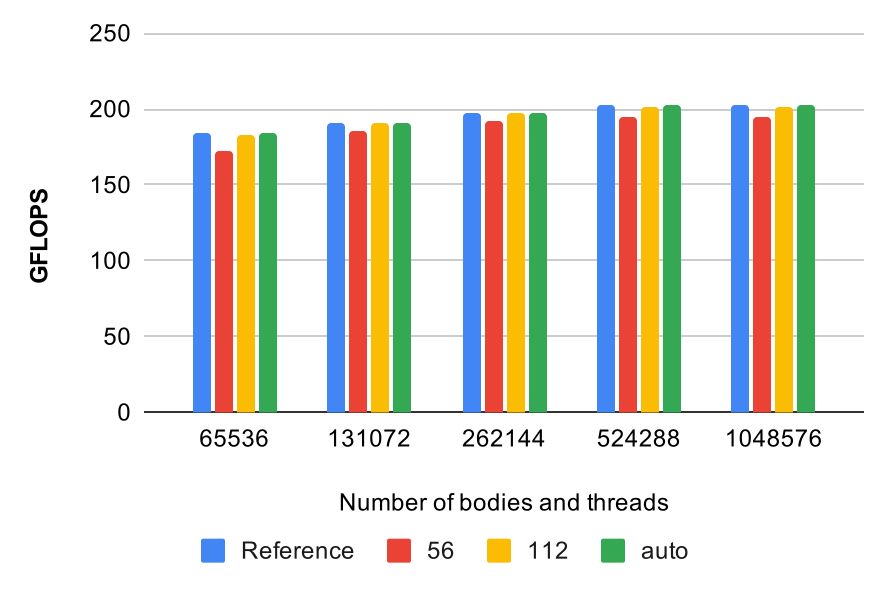}
    \caption{Performance obtained when using the \textit{fold} iterator}
    \label{res_fold}
\end{figure}

Fig~\ref{res_fold} shows the performances obtained when modifying the parallel algorithm using the \textit{fold} iterator. Clearly, Rust does not add any costs when using aspects of high-level languages such as these types of iterators. This allows improving the code to make it more understandable and scalable, without sacrificing perfomance.

\begin{figure}[htbp]
    \centering
    \includegraphics[width=0.95\columnwidth]{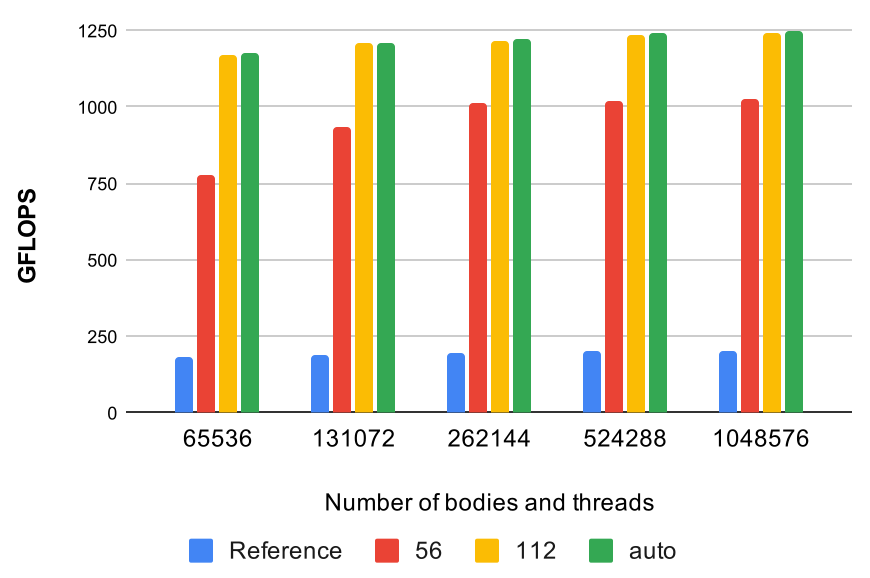}
    \caption{Performance obtained when applying mathematical optimizations}
    \label{res_matematica}
\end{figure}

When applying the mathematical optimizations (Fig~\ref{res_matematica}), the difference in performance versus the previous version is notable. FLOPS rates increase up to 6.1-6.4$\times$ with 112 threads (the best case). However, it is important to remark that this performance benefit comes at the cost of precision relaxation. 

\begin{figure}[htbp]
    \centering
    \includegraphics[width=0.95\columnwidth]{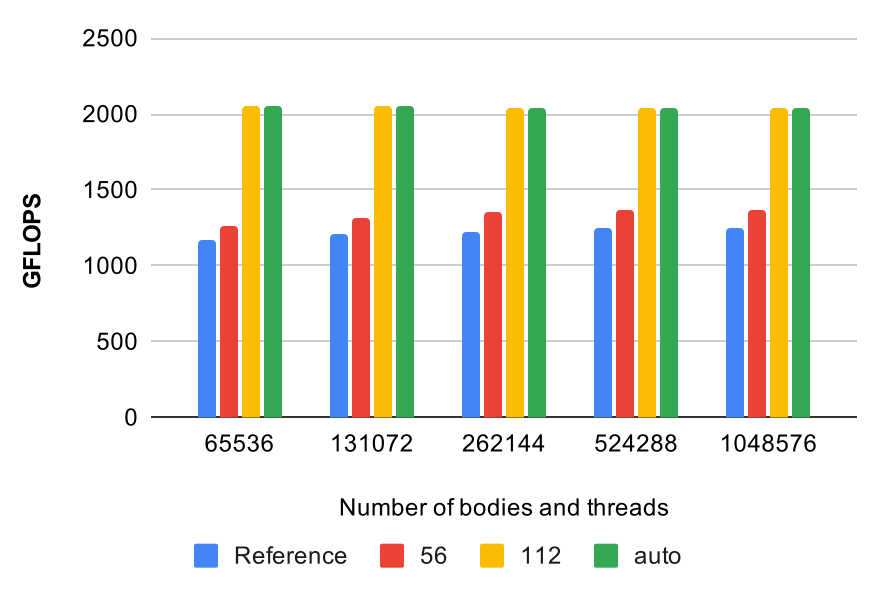}
    \caption{Performances obtained when vectoring with AVX-512 instructions}
    \label{res_vectorizacion}
\end{figure}

Fig~\ref{res_vectorizacion} shows the performances when the compiler is instructed to vectorize the code using AVX-512 instructions. Applying this optimization increased performance by approximately $68\%$. Rust detects the type of hardware and tries to vectorize using the corresponding instructions by default. By explicitly indicating the type to be used, Rust translation can be improved by incorporating specific instructions such as \texttt{vrsqrt14ps} (or its 64-bit variant \texttt{vrsqrt14pd}), used for the reciprocal square root approximation.

\begin{figure}[htbp]
    \centering
    \includegraphics[width=0.95\columnwidth]{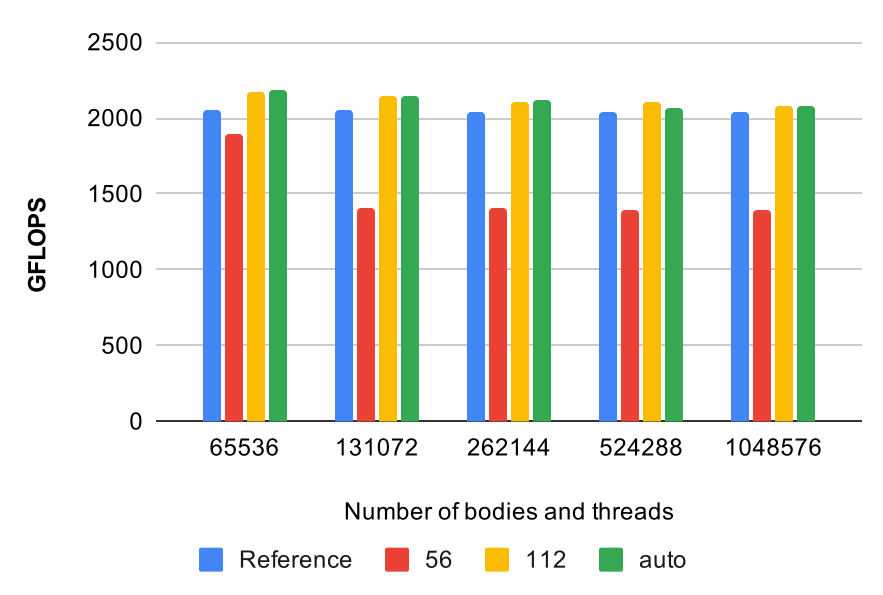}
    \caption{Performance obtained when using Jemalloc as default allocator}
    \label{res_jemalloc}
\end{figure}

Fig~\ref{res_jemalloc} shows the performance obtained when using Jemalloc instead of the memory allocator. This modification produces an improvement of approximately $1.1\times$ when using T=\{112, auto\}. Even though this is not a great improvement, its cost is practically zero, so its inclusion is productive.

\begin{figure}[htbp]
    \centering
    \includegraphics[width=0.95\columnwidth]{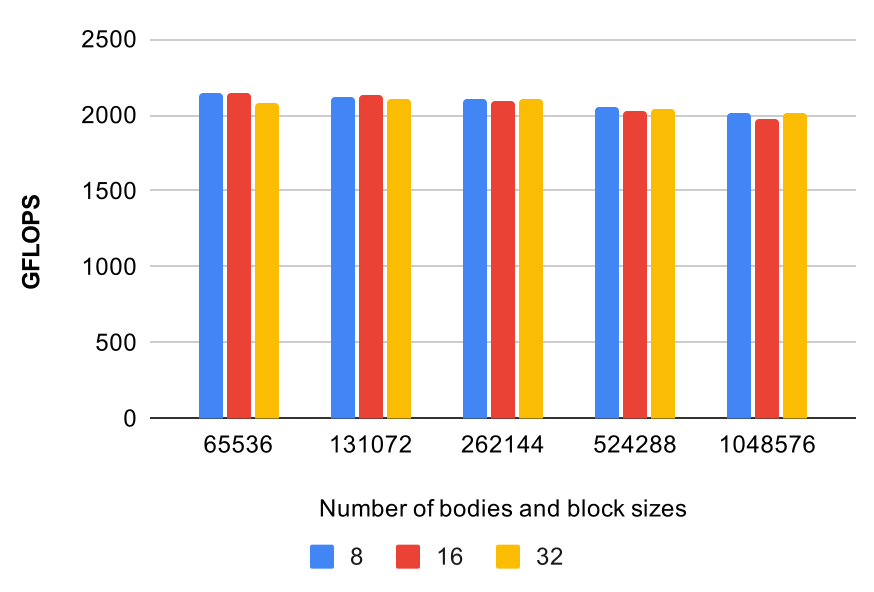}
    \caption{Performance obtained when applying block processing with different block sizes}
    \label{res_bloques}
\end{figure}

Finally, a block processing optimization is applied. Fig~\ref{res_bloques} shows, for each block size and number of bodies, the performance obtained. These measurements were made by setting the number of threads automatically, since this was the configuration with the best performance. As it can be seen, there are no significant differences in the performance obtained with the different block sizes.

\begin{figure}[htbp]
    \centering
    \includegraphics[width=0.95\columnwidth]{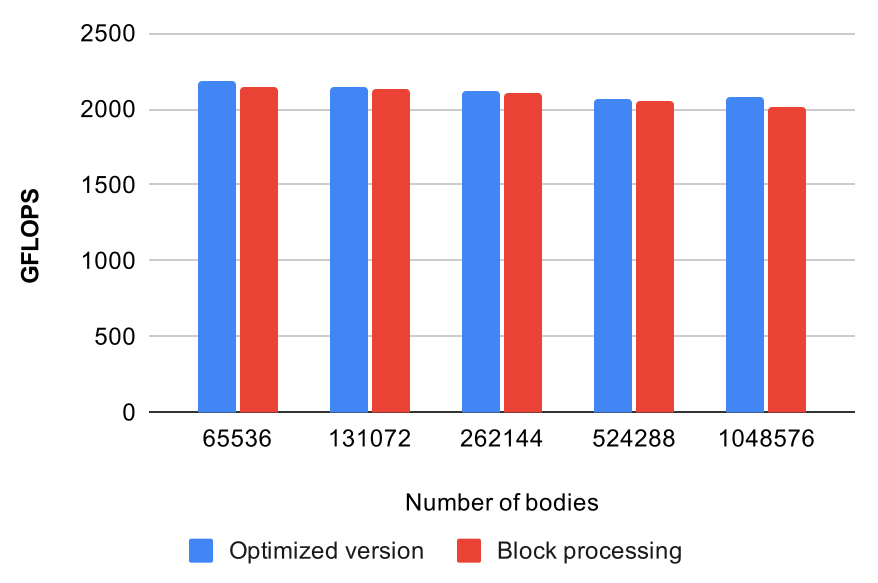}
    \caption{Performance comparison when applying block processing}
    \label{res_rust_final_comp}
\end{figure}

Fig~\ref{res_rust_final_comp} presents a comparison between the last optimized version and the version that processes by blocks. It is clear that block processing does not improve application performance. This is because Rust iterators manage memory efficiently, taking advantage of data locality implicitly.

\begin{figure}[htbp]
    \centering
    \includegraphics[width=0.95\columnwidth]{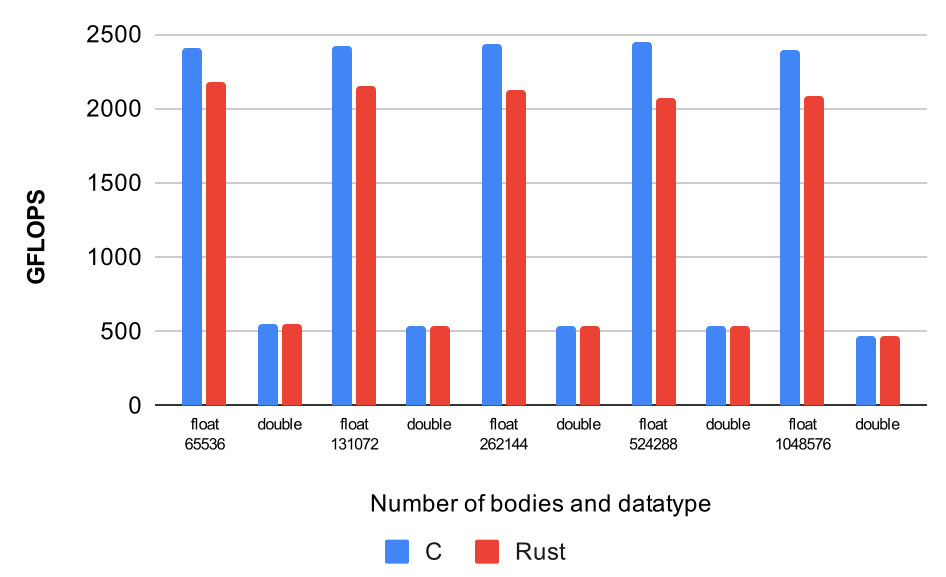}
    \caption{Performance comparison between the most optimized versions of C and Rust}
    \label{res_rust_c}
\end{figure}

Finally, Table~\ref{res_rust_c} shows the final comparison between C and Rust, varying the number of bodies and the data type between single precision (float) and double precision (double). In single precision, the C version outperforms Rust for all problem sizes, achieving improvements of up to $1.18\times$, while in double precision, both implementations have practically the same performance. When analyzing the assembler code generated by both implementations, it can be seen that C performs a more efficient translation of the main code when mathematical optimizations (precision relaxation) are used. This behavior is not replicated in double precision, where both codes are very similar. As already explained, these optimizations are not included in the stable version of Rust yet, so this is expected to improve in the future.

\subsection{Programming Effort}
In this work, not only application performance is of interest, but also the programming effort required for development and maintenance. There are numerous approaches to measuring programming cost, including counting the number of lines of code (\textit{Source Lines Of Code}, or \textit{SLOC}) or the number of characters (including blank lines and comments). Despite their simplicity, these parameters do not reflect algorithm complexity~\cite{sloc}. Other alternatives measure development time, but this is dependent on programmer experience and is not always measurable, as in this work~\footnote{Since an existing version of C has been used, it was not possible to measure this.}.

The subjectivity of these metrics makes it difficult to assess programming cost. For these reasons, in this study, in addition to measuring programming cost through the SLOC indicator, a qualitative comparison of the effort required in both solutions is made. Since both parts are complementary, they offer a broader view to understand the programming effort required.

\begin{table}[htbp]
\begin{center}
\begin{tabular}{|c|c|c|}
\hline
\textbf{} & \textbf{\textit{C}}& \textbf{\textit{Rust}} \\
\hline
Main & 66 & 40  \\
Total & 219 & 195  \\
\hline
\end{tabular}
 \caption{SLOC in final algorithms}
\label{sloc}
\end{center}
\end{table}

In particular, as shown in Table~\ref{sloc}, the main code block of this algorithm in C has \textbf{66} lines of code (Figure~\ref{c_code}), while the Rust version has only \textbf{40}. In total, modularizing the solution, along with all necessary lines, \textbf{219} lines of code were needed in C, versus \textbf{195} in Rust. Considering the main code block, C required $65\%$ more SLOC than Rust. Even though this percentage is reduced by $12.5\%$ when considering the entire code, it should be noted that in Rust, classes were created to represent a force, a body and a position, each of these with its respective methods. In general, this has a favorable impact on program maintainability. Additionally, Rust has the advantage that, being a language with some of the features in high-level languages, both functional and object-oriented, it allows the development of more compact and more self-explanatory code than C.

From a qualitative point of view, the following aspects can be considered: C required changing the logic of the solution to adapt it to block processing in order to take advantage of data locality. This requires both finding the optimal block size, as well as making modifications to the code to ensure proper processing. For this algorithm, Rust efficiently managed memory and no changes to the solution were required. Furthermore, Rust allowed generating parallel code in a simple way (by adding the prefix \textit{\_par} to iterators), while in C, different OpenMP options were required to achieve adequate parallelization. Another important aspect to highlight is that in Rust, external libraries can be easily added, such as the library of mathematical optimizations or Rayon's library. All you have to do is specify the name of the library and its version in a configuration file  (\textit{Cargo.toml}), and it will be available for use in the next compilation.

In favor of C versus Rust, math optimizations can be enabled by sending the \texttt{-fp-model fast=2} and \texttt{-qopt-zmm-usage} flags to the compiler. On the other hand, Rust requires using language intrinsics, which, even though libraries can be used to avoid having to make large modifications to the code, is still an unfavorable point of the language so far.

\section{Related Works}\label{trabrel}

To date, only a few preliminary studies can be mentioned that explore the capabilities of the Rust language for parallel processing applications.

In 2015, a thesis~\cite{wilk15} was published that evaluates the performance and productivity of the Rust, Go and C languages for calculating the shortest path in graphs. As in this work, the Rust version was the one that involved the lowest programming cost.

In~\cite{blanco16}, the authors explore the advantages and disadvantages of Rust in astrophysics by re-implementing fundamental parts of the Mercury-T software (originally written in Fortran). The authors conclude that Rust helps avoid common mistakes both in memory access and race conditions. They also state that the initial learning curve of the language is expensive, but that, once learned, it reduces programming times considerably.

In~\cite{hansen16}, serial N-Body codes are compared considering the Rust and C languages. The authors conclude that they could develop a parallel solution in Rust in a simple way, but that it is still in the process of optimization. This indicates that Rust can produce code that is not efficient, so best practices should be followed when writing code


\section{Conclusions and Future Work}\label{conclusiones}

The choice of programming language is a fundamental decision that will have an impact on system development and performance. To evaluate the feasibility of using Rust in the HPC environment, the N-Body algorithm, which is a problem that requires intensive and parallel computation, was used in this work. To this end, Rust was used to create algorithms, starting with a base version and then, by applying incremental optimizations, a final, improved version was obtained. These optimizations were as follows:

\begin{itemize}
    \item Multi-threading: Given the base code, the parallel version was obtained by simply adding the \textit{\_par} prefix. Additionally, Rust correctly selected the number of threads automatically.
    \item Modification of procedural iterator (\textit{for\_each}) by reduction iterator (\textit{fold}): This modification had almost no impact on performance, but it increased code self-understanding.
    \item Mathematical optimizations: Even though these optimizations caused a loss of precision in the results, they did significantly improve algorithm performance. A library that allows incorporating (almost) implicitly these optimizations without the need to alter the code was used.
    \item Vectorization: Rust was able to auto-vectorize code (i.e, manual vectorization was not required). However, the absence of a vectorization primitive in this language (like \texttt{simd} from OpenMP) can be a limitation when auto-vectorization is not achieved. Additionally, the underlying set of SIMD instructions should be passed to the compiler, since this helps achieving a better selection of instructions. 
    \item Memory allocator: Performance increased when using Jemalloc as the memory allocator, instead of the default allocator. While performance gain was small, its cost was practically zero.
    \item Block processing: This optimization produced no improvements, which avoids having to modify algorithm logic to process the bodies in blocks.
\end{itemize}

After applying the optimizations, the resulting Rust algorithm was compared to its C counterpart. Performance results were close in double precision, but not in single precision, where C version was superior. This is because Rust does not optimize math operations as well as C with this data type.

As regards programming effort, Rust, unlike C, has some high-level language features, which facilitates the generation of code that is easy to maintain. Moreover, since it has features from functional and object-oriented languages, it allows generating more compact code, which resulted in fewer lines of code in the main block of the program.
In addition, Rust tries to manage memory efficiently, and in some cases – like this work – no modifications to the computational logic are required to take advantage of data locality. 

Based on the results obtained and the analysis carried out, we found that Rust offers the benefits of high-level languages, both object-oriented and functional, but without loosing significant performance in these abstractions. Thus, it is considered that Rust can be positioned as an alternative to C for HPC in contexts similar to those of the present study. Since the language is still in constant evolution, community support will become a determining factor in its final viability.

As future work, it would be interesting to extend the study carried out considering three possible aspects:
\begin{itemize}
\item At the top level, selecting other case studies that are computationally intensive but whose characteristics are different from those in the case study selected for this work (f.e. memory-bound applications). 
\item At the middle level, including other programming languages, like C++, would complement the comparison since the latter can be seen as an extension of C but shares some high-level features with Rust.
\item At the low level, considering other HPC architectures, such as AMD processors. 
\end{itemize}
All these extensions would contribute to increase representativeness of this study.

\section*{Acknowledgements}

ER and ADG are members of the Research Career of CIC and CONICET, respectively.

\bibliographystyle{IEEEtran}  
\bibliography{ref.bib}

\end{document}